\documentclass[12pt,preprint]{aastex}
\usepackage{multirow}

\newcommand{\masyr}{\mbox{mas~yr$^{-1}$}}

\newcommand{\GPB}{\mbox{\em{GP-B}}}

\newcommand{\IMP}{\mbox{\objectname[HR 8703]{IM~Peg}}}
\newcommand{\REFA}{\objectname[3C454.3]{3C~454.3}}
\newcommand{\REFB}{\objectname[IAU B2250+194]{B2250+194}}
\newcommand{\REFC}{\objectname[IAU B2252+172]{B2252+172}}
\defcitealias{GPB-I}{Paper~I}
\defcitealias{GPB-II}{Paper~II}
\defcitealias{GPB-III}{Paper~III}
\defcitealias{GPB-IV}{Paper~IV}
\defcitealias{GPB-V}{Paper~V}
\defcitealias{GPB-VI}{Paper~VI}
\defcitealias{GPB-VII}{Paper~VII}

\slugcomment{Accepted to the Astrophysical Journal Supplement Series}

\shortauthors{Lebach et al.}
\shorttitle{VLBI for {\em Gravity Probe~B}.  IV.}

\begin{document}

\title{VLBI for {\em Gravity Probe~B}.  IV.  A New Astrometric Analysis
Technique and a Comparison with Results from Other Techniques}

\author{D. E. Lebach\altaffilmark{1},
N. Bartel\altaffilmark{2},
M.~F.~Bietenholz\altaffilmark{2,3},
R.~M.~Campbell\altaffilmark{4},
D.~Gordon\altaffilmark{5},
J.~I.~Lederman\altaffilmark{2},
J.-F.~Lestrade\altaffilmark{6},
R.~R.~Ransom\altaffilmark{2,7},
M.~I.~Ratner\altaffilmark{1}, and
I.~I.~Shapiro\altaffilmark{1}}

\altaffiltext{1}{Harvard-Smithsonian Center for Astrophysics, 60
Garden Street, Cambridge, MA 02138, USA}

\altaffiltext{2}{Department of Physics and Astronomy, York University,
4700 Keele Street, Toronto, ON M3J~1P3, Canada}

\altaffiltext{3}{Now also at Hartebeesthoek Radio Astronomy Observatory,
PO Box 443, Krugersdorp 1740, South Africa}

\altaffiltext{4}{Joint Institute for VLBI in Europe,
Oude Hoogeveensedijk 4, 7991 PD Dwingeloo, The Netherlands}

\altaffiltext{5}{NVI Inc./NASA Goddard Space Flight Center, Greenbelt, MD 20771}

\altaffiltext{6}{Observatoire de Paris, Centre national de la
recherche scientifique, 77 Av.\ Denfert Rochereau, 75014 Paris,
France}

\altaffiltext{7}{Now at Okanagan College, 583 Duncan Avenue
West, Penticton, B.C., V2A 2K8, Canada and also at the National
Research Council of Canada, Herzberg Institute of Astrophysics,
Dominion Radio Astrophysical Observatory, P.O. Box 248, Penticton,
B.C., V2A 6K3, Canada}


\begin{abstract}
When
VLBI observations are used to determine the position or
motion of a radio source relative to reference sources nearby
on the sky, the astrometric information is
usually obtained
via:
(i)~phase-referenced maps; or (ii)~parametric model fits
to measured fringe phases or multiband delays.
In this paper we describe a ``merged"
analysis technique which combines
some of the most important
advantages
of these other
two
approaches.
In particular, our merged technique combines the superior model-correction
capabilities of parametric model fits with the ability of
phase-referenced maps to yield astrometric measurements of sources that
are too weak to be used in parametric model fits.
We
compare the results from this merged technique with the results
from phase-referenced maps
and
from parametric model fits
in the analysis
of astrometric VLBI observations of the radio-bright star
IM~Pegasi (\objectname[]{HR 8703}) and the radio source
\REFC\ nearby on the sky.
In these studies 
we use central-core components of
radio sources \REFA\ and \REFB\ as
our positional references.
We
obtain astrometric results for \IMP\ with
our merged technique even when the source is too weak to be used
in parametric model fits, and
we find that our merged technique yields superior astrometric
results to the phase-referenced mapping technique.
We used our merged technique to estimate the proper motion and other
astrometric parameters of
\IMP\ in support of the NASA/Stanford {\em Gravity Probe~B} mission.

\end{abstract}

\section{Introduction}
\label{sintro}

Very-long-baseline interferometry (VLBI) provides the most
accurate astrometric measurements of celestial objects currently
attainable.  VLBI astrometry has been used,
among other applications, to: define the most nearly inertial reference
frame
available
for the positions of celestial objects
\citep[e.g.,][]{Ma+1998, FeyGJ2009};
tie an inertial (extragalactic) reference frame to a
planetary ephemeris via observations of pulsars
\citep[e.g.,][]{Bartel+1985, Rodin+2002, Dodson+2003};
characterize
motions and other properties of the Earth
\citep[e.g.,][]{Ryan+1993, Mathews+1995};
study positions and motions of maser spots in galaxies as a means
to estimate black-hole masses
\citep[e.g.,][]{Ishihara+2001, Kondratko+2004}
and distances to other galaxies
\citep{Herrnstein+1999};
and test general relativity via measurements of solar gravitational
deflection \citep[e.g.,][]{Counselman+1974, Shapiro+2004}.
In this paper we describe the analysis of astrometric VLBI observations
of the RS~CVn binary star IM~Pegasi
(IM~Peg; \objectname[]{HR 8703}),
the radio-bright star which served as the ``guide star,'' and hence as the
positional reference,
for the NASA/Stanford {\em Gravity Probe~B} (\GPB\@) experiment.

\GPB\@ was designed to measure the geodetic and frame-dragging effects
predicted by general relativity \citep[see Paper~I,][]{GPB-I},
as manifested in secular changes in the spin-axis orientations of
four gyroscopes
placed within a spacecraft in a low-altitude, polar orbit about the Earth.
The spacecraft also had an on-board telescope
equipped with a tracking system
designed to keep the guide star,
\IMP , at the center of the telescope
field of view, and thereby
provide a directional reference:  the
relativistic effects were estimated by measuring
the drift rates in
the spin-axis directions of the four gyroscopes relative
to the direction to \IMP .
Thus,
if any motions of \IMP\ on the sky relative to
an inertial frame were not
accounted for
sufficiently accurately,
they
could map directly into the gyro drift-rate signals and
thereby corrupt the relativistic measurements.
(Ideally the \GPB\@ guide star would have been
a distant quasar,
which could be treated as effectively motionless on the sky,
rather than a star in our galaxy, but
the on-board telescope could only track objects brighter than about
6th magnitude, so a relatively bright star was the only possible choice.
A list of criteria that had to be met by the guide star,
and a description of the selection process that
led to the choice of \IMP , is in \citetalias{GPB-I}.)
Our VLBI observations of
\IMP\ were therefore undertaken
to determine
the motions of the star relative to an inertial frame
so that
the relativistic effects
included in the gyro measurements could be properly separated.
The goal of our VLBI observations was to determine,
relative to an inertial frame,
the proper motion of \IMP\  
with standard error 0.14~\masyr\  or less in
each of the north-south and east-west directions.
This accuracy goal
was based
upon an error projection for the full \GPB\@ experiment
prior to launch in April 2004 and the desire at the time for the
uncertainty in the proper motion
of \IMP\ not to constitute a significant source of error for the
experiment.

In this paper we present a new approach for astrometric
VLBI data analysis that we developed to meet the
accuracy goal for \GPB\@. 
We then compare the results from this approach with the results
from two well-established VLBI analysis techniques:
phase-referenced mapping (PRM) and ``phase connection'' followed
by parametric model fitting (PMF).
Our new approach is
a combination of these
two standard approaches that provides many of the benefits of each.
All three approaches 
use {\it differential\/}
VLBI astrometry, in which
the target source (i.e., the source of astrometric interest)
is observed alternately with
at least one extragalactic, compact reference source nearby on the sky
to provide
model corrections
in the analysis of
the target-source data.
The way these
corrections are applied differs among the approaches,
but in all cases
the astrometric measurements of the target source are relative to the
position(s) of the reference source(s).

In \S~\ref{vlbiobs} we briefly describe the VLBI observables
used in our analysis.
We then discuss some basic aspects of the PRM and PMF analysis techniques 
in \S~\ref{prmaps} and \S~\ref{pmfit}, respectively.
In \S~\ref{merged} we describe our new ``merged''
analysis technique and then our implementation of it.  We 
compare the results from the three techniques in
\S~\ref{compare}, and offer corresponding conclusions
in \S~\ref{conclusion}.

\section{VLBI Observables: Fringe Phase, Fringe Rate, and Multiband Delay}
\label{vlbiobs}

In VLBI observations of a compact source,
the total fringe phase (or visibility
phase, or interferometric phase, or simply ``phase''),
$\phi(\omega, t)$,
is the phase
at frequency $\omega$ and time $t$
that is
associated with the difference in
arrival times of signals
received
from the source at two antennas in a VLBI array.
We describe $\phi(\omega, t)$ as:
\begin{equation}
\phi(\omega, t) = \omega \left[ \tau_{\rm{geom}}(t)
 + \tau_{\rm{inst}}(\omega, t)
 + \tau_{\rm{atm}}(\omega, t)
 + \tau_{\rm{struc}}(\omega, t) + \tau_{\rm{noise}}(\omega, t) \right]
 + 2\pi N(\omega, t)
\label{phi}   
\end{equation}
where 
$\tau_{\rm{geom}}(t)$, the ``geometric delay,'' is the difference
in the signal arrival times in vacuum at
the two antennas;
$\tau_{\rm{inst}}(\omega, t)$ represents the difference in the
instrumental delays (including clock behavior) at the two antenna sites;
$\tau_{\rm{atm}}(\omega, t)$ represents the
difference in signal propagation times to the two antennas
due to all atmospheric effects,
including the contributions of
the hydrostatic (or ``dry'') atmospheric constituents,
atmospheric water vapor, and the ionosphere;
$\tau_{\rm{struc}}(\omega, t)$ is the delay contribution from source
structure, i.e., from the non-pointlike brightness distribution of the
source;
$\tau_{\rm{noise}}(\omega, t)$ represents the (thermal) noise contribution 
to the phase measurement;
and $N(\omega, t)$ represents the integer number of ``ambiguities,'' or
``phase wraps,'' included in the measurement.
Information about the number of phase wraps associated with
a fringe phase is not inherent to a measurement of fringe phase,
and in general a measured fringe phase can be defined such that
$-\pi < \phi_{\rm meas}(\omega, t) \leq \pi$, or
$0 \leq \phi_{\rm meas}(\omega, t) < 2\pi$.
There is a fringe phase associated with each antenna pair, or
``baseline,'' within the VLBI array.
The ``fringe rate'' is the partial derivative of the fringe phase
with respect to time.
The ``multiband delay'' is a measured approximation
of the group delay, which is the partial derivative of the fringe phase with
respect to frequency.

To lowest order (i.e., neglecting, e.g., relativistic effects),
the geometric delay is given by:
\begin{equation}
\tau_{\rm{geom}}(t) = {1 \over c}\left[ {\mathbf B}(t) \cdot {\mathbf{\hat s}}(t)\right]
\label{geom} 
\end{equation}
where
$c$ is the speed of light in vacuum,
${\mathbf B}(t)$ is the 3-dimensional vector between the two antennas of the
baseline, and
${\mathbf{\hat s}}(t)$ is the unit vector in the direction of the
observed source.
All of the astrometric information in the measured fringe phase is
contained in $\tau_{\rm{geom}}(t)$.
(While we do not show
the relativistic contributions to $\tau_{\rm{geom}}(t)$ in Equation~\ref{geom},
we do include those contributions throughout our analyses.
We also account for
Earth motions relative to the solar-system barycenter
during the time between
signal arrival at the two antennas.)

The model we used for
$\tau_{\rm{atm}}(\omega, t)$
can be more explicitly described by:
\begin{equation}
\tau_{\rm atm_A}(\omega, t, \epsilon, \mathbf{\hat s}) =
   \tau_{\rm zen\_dry_A}(t)m_{\rm dry_A}(t, \epsilon)
 + \tau_{\rm zen\_wet_A}(t)m_{\rm wet_A}(t, \epsilon)
 + \tau_{\rm ion_A}(\omega, t, \mathbf{\hat s})
\label{tau_atm}   
\end{equation}
where the
``$A$'' subscript refers to site A;
$\tau_{\rm zen\_dry_A}(t)$ is the propagation delay through the
atmosphere at zenith under the assumption that the atmosphere is
in hydrostatic equilibrium;
$\tau_{\rm zen\_wet_A}(t)$ is the additional propagation delay through the
atmosphere at zenith due to tropospheric water vapor
\citep{Davis+1985}; 
$m_{\rm dry_A}(t, \epsilon)$ and $m_{\rm wet_A}(t, \epsilon)$ are, respectively,
the ``mapping functions'' that project (i.e., scale)
the ``dry'' and ``wet'' delays at zenith to the
line-of-site elevation, $\epsilon$, of the observed source;
and
$\tau_{\rm ion_A}(\omega, t, \mathbf{\hat s})$ 
denotes the line-of-site
contribution of the ionosphere to phase delays
in the direction of the observed source.

\section{Phase-Referenced Mapping (PRM)}
\label{prmaps}

The basic
idea behind
the PRM technique
is to use the data from
the observed reference sources to account for otherwise unmodeled fluctuations
in instrumental or atmospheric delays, as well as for other model errors.
Specifically,
the ``residual'' fringe phases and rates, i.e., the differences between
the measured and a~priori model values of these quantities,
are obtained for the reference sources and then
temporally and sometimes also spatially
interpolated to the observation time and sky position of the target source
to estimate the effects of
model errors
on the target-source observables
\citep[see, e.g.,][]{Shapiro+1979,
Gorenstein+1983, Lestrade+1990, BeasleyC1995, Fomalont+2005b}.
Typically the cycle time over which the
reference sources and target source are observed is relatively
short, from several seconds to several minutes.
To the extent that reference-source structure is properly accounted for
and the reference-source
residuals are properly interpolated to the 
observation time and sky position of the target source,
the remaining residual components in the
target-source data can be attributed primarily to
measurement noise,
unmodeled target-source
structure, and a position offset of the target source
relative to the a priori model position.
Perhaps the most important feature of the PRM technique
is that the
target-source data can be coherently integrated over the entire
span of an observing session.
Thus, even very weak sources with
flux densities well under 1 mJy can be detected and imaged with
VLBI via this technique.

Henceforth in this paper,
a ``scan'' refers to a single continuous observation of a particular
source.
Fringe-phase and fringe-rate measurements from successive reference-source
scans
can provide proper model fringe-phase
adjustments for intervening target-source data only if
the change in the model errors between the scans is $\ll 2\pi$,
so that
the number of $2\pi$ phase wraps between
successive reference-source scans can be accurately tracked.
For that reason, the cycle time over which the reference and target sources
are observed should be as short as possible.  On the other hand,
the reference-source scans must
be long enough
to provide reliable fringe-phase measurements from single scans, and the scan
times for the target source should be
long enough to ensure
sufficient coherent integration time over the course of the
observing session to produce an image with acceptably high
signal-to-noise ratio (SNR).
These trade-offs must be balanced to determine the ``optimum'' scan and cycle
times in the observation schedule.

Another beneficial feature of the PRM approach is that
it is relatively quick and efficient  for obtaining
high-accuracy astrometry, in large part due to the software
packages now readily available.  In our analyses we used
almost exclusively the Astronomical Image Processing System (AIPS)
provided by the National Radio Astronomy Observatory (NRAO) to produce
our phase-referenced maps.
We followed the guidelines of \citet{Diamond+1995}.
Specifically, we:

\begin{enumerate}
\item Incorporated the instrumental phase and amplitude calibrations
routinely provided
in log files from individual antennas.  We also unweighted faulty
data based upon information in
these files and from operator and correlator reports.

\item Incorporated an additional constant (over time) phase adjustment
to the calibration for each antenna
based upon ``fringe fitting'' (via AIPS task FRING) of data
from a selected reference scan.\footnote[1]{The use of such constant phase
adjustments based on data from selected scans is referred to as
``manual'' phase calibration \citep{Diamond+1995}.
Even when measured phase calibrations were available in log files to account
for instrumental effects,
we found that
we obtained better calibration
across our observed bandwidth when we used manual phase calibrations
after the application of our measured phase calibrations.
This finding was based upon
an assessment of the phase scatter across the spanned bandwidth of data
from some of our 1997 and 1998 observations.
As a result,
we used manual phase calibrations for all data after
we applied any available measured phase calibrations.}
Sometimes we used multiple reference
scans to provide such phase calibration adjustments for
all antennas.

\item Ran a ``global fringe fit'' (via AIPS task FRING) on all
reference-source
data while applying the phase and amplitude calibrations obtained so far.

\item Ran AIPS task IMAGR, 
with the fits from Step 3 as calibration,
to generate ``self-cal'' maps of the
reference sources
\citep[see Paper~II,][]{GPB-II}.   We then used
these self-cal maps
to refine the amplitude and phase calibration as well as the data flagging.
In general we repeated Steps 3 and 4 several times.

\item Used AIPS task BPASS to further
refine the phase and amplitude calibrations as a function of frequency.

\item Used AIPS task IMAGR with the final calibrations from Steps 1
through 5 to obtain a phase-referenced map of the target source.

\end{enumerate}

Unfortunately, the use of the PRM technique for
astrometry also poses some challenges, especially for campaigns such
as ours in which the observations are made over many years.
For example, one must take great care to assure that model components
such as site positions, antenna axis offsets, and Earth orientation
parameters (EOPs, i.e.,
X- and Y- pole positions, UT1$-$UTC, and nutation angles in
longitude and obliquity; see, e.g., \citealt{Seidelmann+1982, Seidelmann+1992})
are modeled
consistently and correctly throughout all observation sessions.
(For example, if updated values
of axis offsets for antennas
are used in processing data from later observing sessions,
then the new values will have to be incorporated into a re-analysis of
data from earlier observing sessions
to avoid possible systematic errors.)
Furthermore, post-processing
adjustments to model parameters can be quite cumbersome in some
analysis packages, including AIPS, especially if the adjustments
have to be made to a large number of experiments.
In addition, without the use of multiple reference sources and
special interpolation routines to handle the spatial (in particular,
elevation-angle) dependencies of atmospheric delays,
a simple temporal interpolation of reference-source fringe phases introduces
model-correction errors that scale roughly with
the angular separation between the reference source and the target
source.
Thus the basic PRM technique
commonly loses viability for an
angular separation between reference and target sources larger
than a few degrees.
Finally,
the PRM technique offers no inherently good way to assess
the effects of systematic errors (such as those due to inaccurate modeling
of the atmosphere).
Reliable estimates of the true accuracy of the astrometric
results can therefore be difficult to obtain
(although see \citealt{PradelCL2006} for a comprehensive assessment
specifically for the VLBA and EVN arrays).
Our
development and
use of the ``merged'' technique described in \S~\ref{merged}
of this paper was motivated in part by our
desire to overcome these drawbacks.

\section{Parametric Model Fitting (PMF)}
\label{pmfit}

The PMF technique for analysis of VLBI data is,
in all implementations with which we are familiar,
essentially the method of
``weighted-least-squares''
estimation.  For each reference-source and
target-source scan, the total measured fringe phase
and fringe rate are estimated for each baseline
at an epoch near the center of the scan period.  Any
$2\pi$ ``phase jumps'' between successive phase measurements
are then resolved (or flagged as unresolvable) for each source and baseline.
The overall integer number
of $2\pi$ differences between the phases from different
sources is also resolved.
Finally, the complete collection of ``phase-connected'' data
is fit to a
model in which corrections to a
wide range of parameters---including instrumental delays,
propagation delays through the atmosphere, and source
positions---are simultaneously estimated.

To avoid the often onerous task of determining the change in the number
of $2\pi$ phase
wraps between successive scans prior to model fitting,
multiband delays
\citep[e.g.,][]{Clark+1985} are commonly used
in place of fringe phases as the principal observables
with the PMF technique.
The use of multiband delays comes at a significant cost in statistical
(i.e., SNR-derived) measurement accuracy relative to the use of fringe
phases, but the use of only bright radio sources as well as
observations over wide spectral bandwidths can reduce this cost.
The PMF approach
with multiband delays
is used by the geodetic VLBI community
in virtually all of their data analyses, including their studies to define
the International Celestial Reference Frame
and track apparent motions of extragalactic sources
\citep{Ma+1998, FeyGJ2009}.
In general, the radio sources chosen for such studies
are the brightest and most compact known.

The PMF technique has several advantages over the PRM technique.
For example, since
total measured fringe phases and rates
(rather than residual fringe phases and rates)
are processed in the analysis, one can make
improvements or changes to a priori models used with multi-year
data sets much more easily.
Furthermore, the use of data from multiple reference sources
to account for spatial (e.g., elevation-angle-dependent)
as well as temporal phase variations is straightforward
and readily implemented with existing software packages.
In addition, because the PMF technique is a method of
weighted-least-squares estimation of parameters that are treated as
Gaussian random variables, one can readily obtain a measure of
the sensitivity of
one
estimated parameter
(such as a source declination) to the variation
in another (such as the propagation delay
through a site atmosphere) by evaluating the covariance matrix
obtained with the parameter estimates.  The effects on estimated
parameters of other model
changes are also straightforward to evaluate, so various sensitivity studies
to assess systematic errors can be readily implemented.

Unfortunately, the PMF technique also has its limitations and
problems.  Unlike the PRM technique, the PMF
technique can make use of only those data for which fringe phases
for a baseline
are reliably detected in a single scan.  Thus the technique is
only well suited for relatively bright sources.  Furthermore, the PMF
technique offers no direct mechanism to account for source
structure.  One can assess fringe amplitudes and ``closure'' phases
\citep{Rogers+1974} to identify the presence of significant structure,
but the model corrections for structure have to be generated
in a separate process.  Finally, when accuracy requirements dictate
that the PMF technique be used with fringe phases
rather than multiband delays,
the
integral number of $2\pi$ phase wraps
between neighboring scans
must be determined, which can be a very labor-intensive process
and sometimes is not even possible.

\section{New Merged Analysis (MA) Technique}
\label{merged}

\subsection{Motivation}

We desired the most accurate astrometry that we could attain
from our VLBI observations of \IMP .
Unfortunately, neither the PRM nor the PMF
analysis technique adequately met our needs.

The PRM technique, as implemented
with AIPS, relied upon insufficiently accurate models
for our astrometric demands,
in particular for
$\tau_{\rm{atm}}(\omega, t)$
and
also for EOPs and other parameters.
These models
were difficult to reliably correct properly within AIPS.
Also, the use of correct, consistent
values for
some model parameters, such as antenna coordinates and axis
offsets,
would have been
somewhat burdensome
to implement and assure
over the 
8.5~year span of our VLBI observations (which are
described further in \S~\ref{obs}).
In addition, we wanted to make full
use of observations of a second reference source to model the effects
of atmospheric gradients on our measured fringe phases, and we had no
satisfactory software tools with this capacity available to us with
the PRM technique.\footnote[2]{In our implementation of the PRM technique,
we used the data
from only a single reference source, \REFA, for phase calibration.
Thus
we had no way to identify phase adjustments that had elevation-angle
dependencies from those that did not.
However,
the PRM technique is not intrinsically limited to the
use of one reference source for phase calibration.  In fact,
software
within the NRAO's AIPS package
\citep[AIPS task ATMCA;][]{Fomalont+2005a}
is
specifically intended to make use of multiple reference
sources with the PRM technique.
This software
became available only at the end of our VLBI campaign,
and
due to time and budgetary constraints, we were unable to use it.
The ability of this software to improve astrometric
accuracy, relative to when a single reference source is used,
has
been demonstrated
\citep{Fomalont+2005b}.}
Finally our use of the PRM technique
made difficult a robust assessment of
several possible sources of systematic error in our astrometric results,
including, for example,
our sensitivity to errors in our a~priori EOP values
or in the various components
of our atmospheric model.

We also could not use the PMF technique
with all of our data,
because during several observing sessions
the radio emissions from \IMP\ were too weak ($\lesssim 1$~mJy) for the
star to be detected in any single scan.  In several other
sessions, the star was detected only intermittently in single scans
and
in those cases
only
on the most sensitive baselines.  Furthermore, our principal reference
source, \REFA, which we selected because
of its close proximity on the sky to
\IMP\ (0.7\arcdeg\ angular
separation) and very high radio brightness, has a complex and evolving
structure \citepalias[see][]{GPB-II} that made necessary
the inclusion of structure corrections in our model.
The PMF technique cannot provide such
corrections,
so if we had relied upon the PMF technique for our astrometry,
then we still would have needed
to image
\REFA\ separately and generate structure models for it.
Likewise, \IMP\ had
highly time-variable and sometimes complex structure
(see Paper VII, \citealp{GPB-VII}, and \citealp{Lebach+1999}),
so we would have had to
image this source separately, too.

Our solution to overcome the drawbacks associated with
each of the PRM and PMF techniques
was to merge
the two approaches in a way that gave us the advantages and shed
many of the disadvantages
of both.

\subsection{Basic Description}
The basic approach behind the MA technique is
first to follow the approach of the PMF technique
to obtain model corrections.
One bases these corrections
on data from only those radio sources---usually
the reference sources---whose emissions are sufficiently bright
that they can be reliably detected in a single scan on many baselines
and hence are well suited for analysis with the PMF technique.
For improved accuracy,
structure corrections for these sources
can be obtained separately and included in the analysis.
The data from the sources---usually the target sources---that
are not well suited for the PMF technique
are left unweighted but used
as ``placeholders''
so that
model corrections corresponding to the
observation times
and positions of those sources are readily
attainable via interpolation of the model results obtained from
the data that are weighted.
In effect, more accurate models for the target-source observables are
obtained than with the conventional PRM technique, because one has
full use of the PMF-technique tools and models that are (in general)
superior to those available with the PRM technique.
The corrected models from the PMF technique are then used
with the PRM technique, in place of the models and corrections that would
have otherwise been used in a conventional implementation
of the PRM technique.
Our MA technique thus
combines the superior
model-correction capabilities of the PMF technique with the superior
sensitivity of the PRM technique.

\subsection{Implementation}
Here we provide details about the
way that we implemented the
MA technique to analyze the \IMP\ data obtained for \GPB\@.

\subsubsection{Observations}
\label{obs}
Our observation strategy, as well as a list of the VLBI antennas we used,
is provided
in \citetalias{GPB-II}.  We repeat
some relevant points about our observations here.

We had 35~sessions of VLBI observations of \IMP\ between
January 1997 and July 2005.
We made all of these observations over a continuum near
8.4~GHz ($\lambda = 3.6$~cm).
Typically we used
data from
12 to 14 VLBI antennas located around the world.
Most of the antennas we used
are within the United States.

\begin{figure}
\centering
\includegraphics[trim= 0in 2.6in 0in 0in, clip, width=\textwidth]{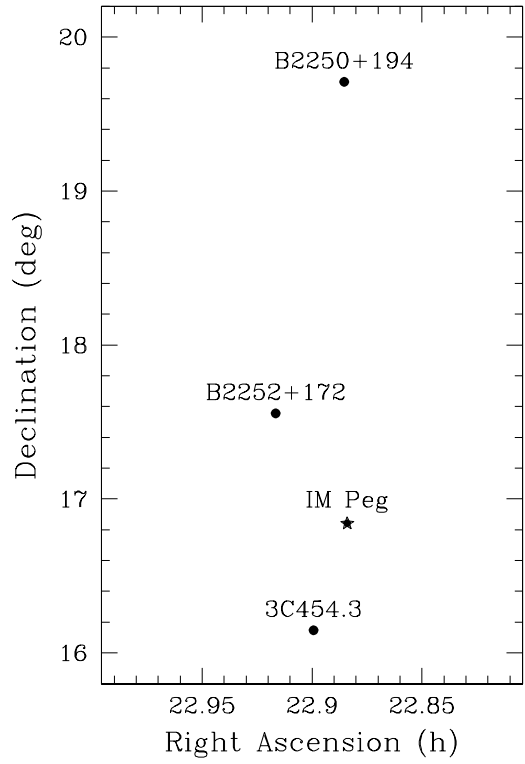}
\caption{Positions (J2000) on the sky of the four radio sources used for
\GPB\@ astrometry.  The east-west and north-south directions on the
plot are shown to the same scale.}
\label{figskypos}
\end{figure}

We sequenced
our observations through either
three or four sources in a repeating 5.5--7~minute
cycle.
The three sources we observed in all 35~sessions were:
our target source, \IMP ;
our principal extragalactic reference source, \REFA\ (B2251+158);
and a secondary extragalactic reference source,
\REFB . 
We added the fourth source, \REFC , to our cycle
for the final 12~sessions.
We treated this source as a second target source and used it
to place bounds on the sizes of apparent changes in the
positions of our reference sources due to structural evolution
(i.e., changes in the brightness distributions of the reference sources).
We also used it to test our different astrometric analysis techniques:
We expected \REFC\ (unlike \IMP ) to be a stationary (extragalactic)
object on the sky, so any estimated motions of the source would be a
measure of the experimental errors inherent to our astrometric
technique.  Figure~\ref{figskypos} shows the relative positions on the
sky of the four sources we observed.  Note that \IMP\ is
between, and approximately collinear with, the reference sources
\REFA\ and \REFB .  We intentionally selected our reference sources
with such alignment
to obtain more accurate
models for \IMP\ data
from the
interpolation of the models fit to our 
reference-source data:
These models account approximately for the
elevation-angle dependencies of our measured fringe phases.
We selected \REFC\ primarily on the basis of its close proximity on the
sky to \REFA\ and secondarily for its approximate alignment, too, between
\REFA\ and \REFB.

The cycle time, and
the scan time on each source per cycle, were chosen with the intent
that they would be: (i) long enough to
get reliable fringe-phase measurements for both reference sources
on all baselines
within a single scan;
(ii) long enough to perhaps get reliable fringe-phase measurements
of the guide star on at least the most sensitive baselines within a single
scan, thereby allowing use of the PMF technique;
and
(iii) short enough to determine
(without ambiguity)
the integral number of
2$\pi$ phase wraps between successive
measurements of each of the sources.
We used two reference sources rather than one primarily so that we could
distinguish model errors that have elevation-angle dependence
(e.g., atmospheric delays) from those that do not (e.g.,
station clock behavior) and thereby improve our overall
astrometric accuracy \citep[see, too,][]{Fomalont+2005b}.

We also analyzed data from four sessions of VLBI observations of
\IMP\ made
by one of us and other colleagues \citep{Lestrade+1999}
between December 1991 and July 1994 in
support of the {\em Hipparcos} mission \citep{Lestrade+1995}.
However,
these four sets of observations
had significant differences from
the later 35~sessions of observations made specifically for
\GPB\@.
Perhaps most significantly, these earlier observations used only four antennas
per session and a single reference source, \REFA.
Therefore we did not include the results from these four earlier sessions
in the comparisons of the astrometric techniques that we
present in this paper.

\subsubsection{First-stage data reduction: calibration and fringe fitting}

We obtained the
amplitude and phase calibration of our data
for the MA technique by following steps 1--5
outlined in \S~\ref{prmaps} for the PRM technique.
We then used AIPS task FRING to estimate
the fringe phases, fringe
rates, and group delays
from the visibility (i.e., cross-correlation) data output from the correlator.
We obtained these estimates on a scan-by-scan basis.
We also obtained these estimates one baseline at a time
by specifying two antennas at a time with the ANTENNA parameter of
FRING.\footnote[3]{AIPS task BLING can also be used to estimate
the observables one baseline at a time. However, technical difficulties
with BLING in the earliest days of our AIPS processing of the experimental
data compelled us
to use FRING instead, and we then opted to continue to use FRING for
all experimental data.}
The visibility data from which these estimates for each baseline were derived
were independent, and thus we followed the common (and greatly simplifying)
practice of treating the estimates for each baseline as independent.

To account for the non-pointlike brightness distribution (i.e.,
structure) of our reference sources, we made images of the sources
using the self-calibration (or ``hybrid mapping'') scheme described
by \citet{Walker+1999}.   We generated one such image for each
reference source
for each session of observations.
We obtained structure corrections for our
estimated fringe phases, fringe rates, and group delays
by running AIPS task FRING twice for each baseline: once with the
self-calibrated CLEAN map used as input calibration, and once with
the source modeled as a point source (which is the FRING default).
We took the structure corrections for the observables to
be the differences between the estimates
with a point-source map and
with the CLEAN map.
An important consideration in the
generation of the structure corrections is the choice
within the CLEAN map
of the reference
position that is defined to have zero structure correction
(i.e., the effective point-source position of the source).
Initially, we used the brightness
peak of the image of each reference source
as the reference position,
because the brightness peak was always clearly defined
and easily identifiable.  However, as we discuss in Paper~III
(\citet{GPB-III}; see also Paper~II),
further studies of the collection of images of our primary reference
source \REFA\ revealed that a different point
within the source
made a substantially better choice
of presumed-stationary reference position.
We thus adopted this point, identified as ``C1'' in
Papers II and III,
as the reference position for our \REFA\ structure corrections.
A set of images
that show the location of ``C1'' within the brightness
distribution of \REFA\ can be found
in \citetalias{GPB-III}.

The
results
from AIPS task FRING for fringe phase, fringe rate, and group delay
are estimates of residuals to the model used by the
correlator.  We used a customized version of AIPS task CL2HF\footnote[4]{
Our customized version of this task
performs the same basic functions.  It includes features
we required (and added ourselves) that were not available
in the earliest versions of CL2HF.}
to add these residual estimates to the
a~priori model values from the correlator
and place the resultant
``total'' observables into AIPS tables.
We then ``exported'' these total observables from AIPS with AIPS
task HF2SV.  The resultant directory of files of total observables was
combined with files of
EOPs,
surface meteorological data, antenna and source coordinates, and additional
calibration data (see below)
to produce databases that we used for the next stage of data
reduction.
We created these databases with the support of the
VLBI group of the Space Geodesy Program at NASA's 
Goddard Space Flight Center.  The theoretical observables and
partial derivatives with respect to model parameters were computed with
that VLBI group's CALC
software
\citep{Caprette+1990}.  
We used CALC version 9.13 and the Jet Propulsion Laboratory's DE200
ephemeris
\citep{Standish1982, Standish1990}
for all of our a~priori models.

Along with the observable data, we also exported from AIPS a
collection of files of the
model fringe phases and rates used by the correlator.  In
the final stage
of our data reduction, discussed
in \S~\ref{mergfin}
below,
we effectively replaced these model phases and rates with improved
models that we derived from
the observable data that we exported.

\subsubsection{Second-stage data reduction: phase connection and parametric
model fitting}

The next step of our data reduction was to use the
measured phases from the reference sources
to improve the
model estimates of the phases of the target source.  In particular,
we used the measured phases from our two
main reference sources,
\REFA\ and \REFB, to separate
a~priori model errors that
have an elevation-angle dependence
(as would be contained in $\tau_{\rm{atm}}(\omega, t)$ in Equation~\ref{phi})
from those that do not
(as would be contained in $\tau_{\rm{inst}}(\omega, t)$ in Equation~\ref{phi}).
We also used these measured phases to identify time intervals
when the phases cannot be well
tracked by existing models.  We
unweighted the target-source data within these intervals, which could
be as short as a single scan of target-source data.

\begin{figure}
\centering
\includegraphics[trim= 0in 3.3in 0in 0in, clip, width=\textwidth]{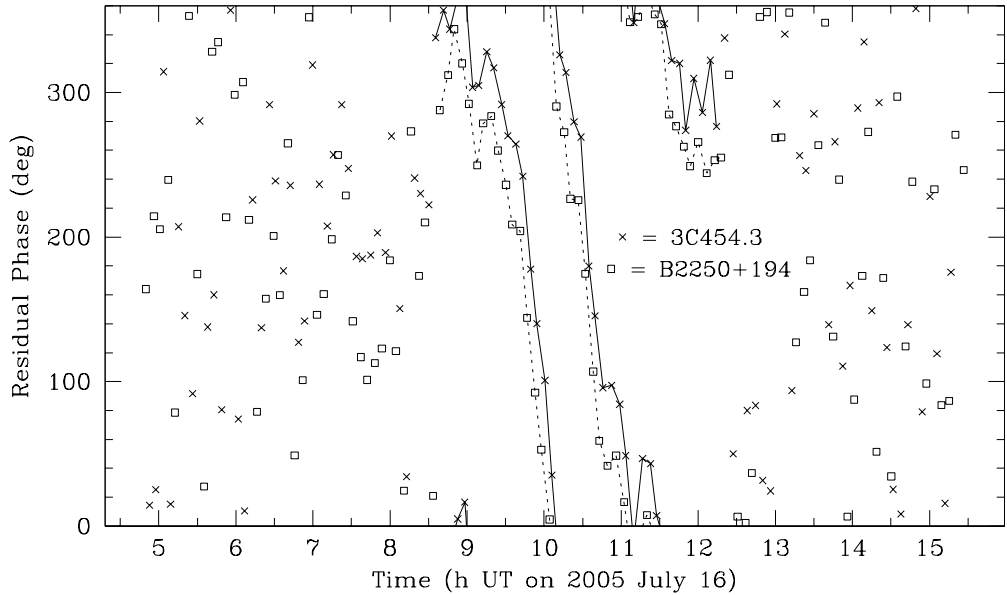}
\caption{Representative phase residuals at the start of the
phase-connection process.  Shown are the differences between
the measured fringe phases and the a~priori model phases for the
Pie Town (New Mexico) to North Liberty (Iowa) baseline of
VLBI observations made on 2005 July~16.
Here
the integral number of phase wraps for each
measurement was selected such that the residual phase is always between
0\arcdeg\ and 360\arcdeg\ (i.e., between 0 and $2\pi$ radians).
We plot only the data for
our two principal reference sources, \protect\REFA\ and \protect\REFB .
Error bars
are not shown to improve plot clarity, but in general
are $\ll 1$ radian.
The section of data with solid (for
\protect\REFA ) and dotted (for \protect\REFB )
lines through the points
corresponds to the period when changes in the model phase errors between
successive measurements were $\ll 2\pi$ radians for both sources, so that
the integral number of $2\pi$ phase wraps between the
measurements could be reliably determined.  Only this ``lined'' section of
the data is considered ``phase connected'' at this stage of the
processing.}
\label{figphcon1}
\end{figure}

\begin{figure}
\centering
\includegraphics[trim= 0in 3.3in 0in 0in, clip, width=\textwidth]{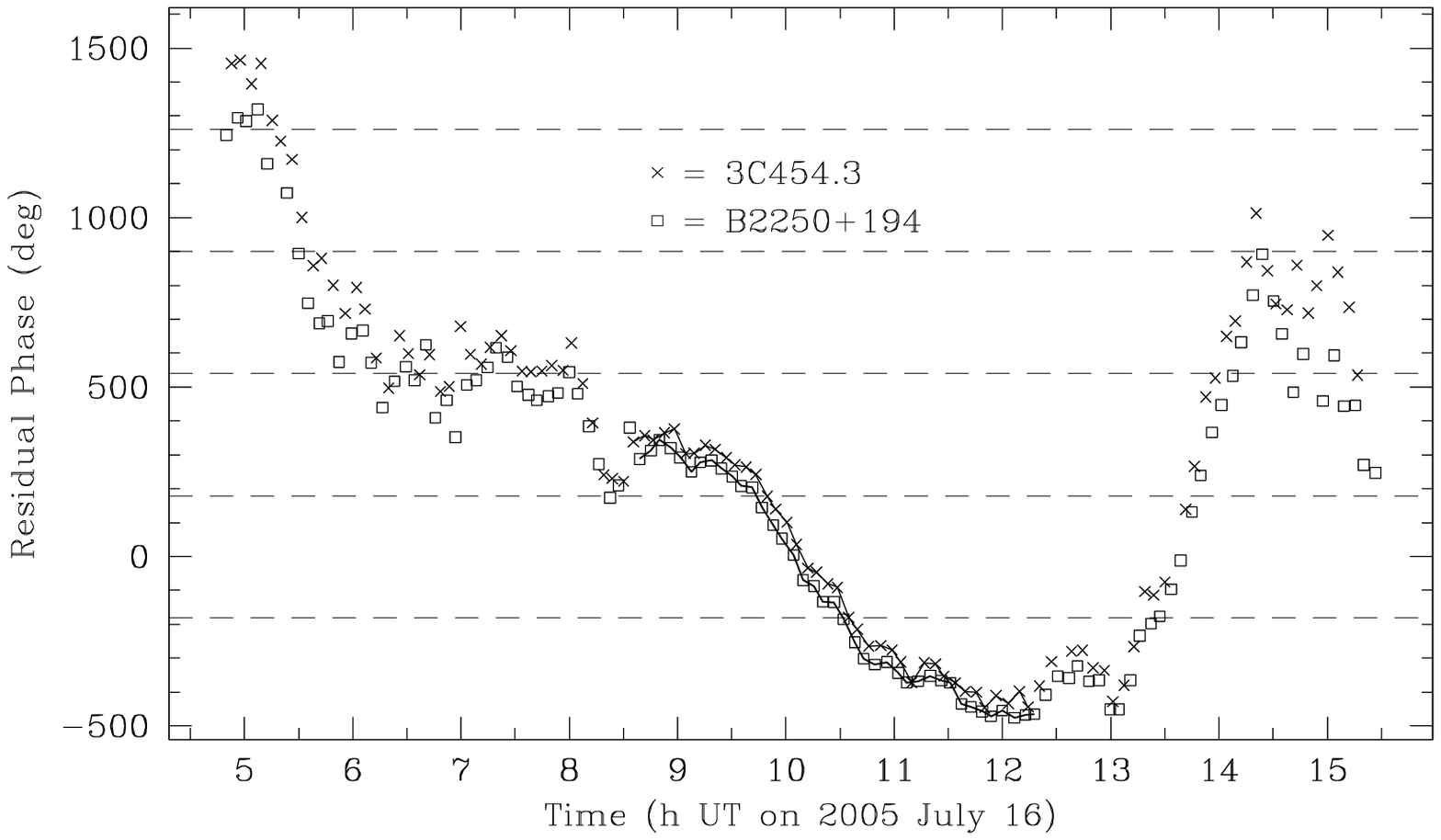}
\caption{The same residual phases as shown in Figure~\ref{figphcon1}
for the Pie Town (New Mexico) to North Liberty (Iowa) baseline
after the integer number of 360\arcdeg\ (or $2\pi$ radian)
phase wraps
were adjusted to remove the obvious $2\pi$ radian phase jumps.
The successive dashed horizontal lines are separated by
$2\pi$ radians.
The data are considered to be reliably phase
connected when the changes in the residual phases between
successive measurements are $\ll 2\pi$ radians.
Only the data points with solid lines through them (one line for the
\protect\REFA\ data and one line for the \protect\REFB\ data) are
assumed to be reliably phase connected at this stage.
The phase-connected residuals still span a range $> 2\pi$ radians;
this result is due to inaccuracies in the preceding model fit that
a subsequent model fit would, in general, remove.
The newly phase-connected
data are used in a subsequent model fit
to estimate corrections to the previous model and
thereby facilitate the phase connection of additional data.  This
process can be iterated until no more $2\pi$ phase wraps
yield residual phase changes $\ll 2\pi$ radians between successive measurements.
}
\label{figphcon2}
\end{figure}

``Phase connection'' refers to the determination of the correct
integral number
of $2\pi$ phase wraps between successive measurements of fringe
phase for a particular baseline and source
(see Figure~\ref{figphcon1}).
Phase connection is possible in part because of
our very accurate a priori models 
and in part because of
the very stable time reference (hydrogen-maser frequency standard) 
used at each VLBI antenna.
Our basic approach to phase connection was to start with the
baselines
and
time segments
of \REFA\  and \REFB\  data for which our
a~priori models were sufficiently accurate that
the number of $2\pi$ phase wraps between successive fringe-phase
measurements was immediately evident
(see Figure~\ref{figphcon2}).
We then used this subset of
phase-connected data to estimate adjustments 
to the parameters of our a~priori model.
The resulting improvement in our model enabled us to phase connect
more data, and so the process proceeded in an iterative manner until
we could no longer reliably phase connect any additional data.
For observations above 10\arcdeg\ elevation, 
we reliably connected, in total, 83\% of the \REFA\ and \REFB\ phases
that we evaluated in this data-reduction stage.

The terms
in Equation~\ref{phi}
that can be associated with difficulties
in the phase-connection process are those for which the model errors can
result in relatively large ($\gtrsim 1$~radian) residual phase fluctuations
in a random pattern over successive reference-source
scans, i.e., in just a few minutes of time.  Our a~priori
values for parameters such as site coordinates and EOPs were sufficiently
accurate\footnote[5]{
Our a~priori values for antenna coordinates came primarily
from the ITRF2000 solution
\citep{Altamimi+2002}
provided by the International Earth Rotation Service
(IERS).
Our a~priori values for the EOPs
also all came from the IERS.}
that they could not cause such rapid residual phase
fluctuations.  In fact, nearly all such phase fluctuations could be
attributed
to changes
either
in
instrumental delays,
$\tau_{\rm{inst}}(\omega, t)$,
which include
possible ``clock jumps'' in the frequency standards at
the sites, or, most commonly, in propagation delays through the atmosphere,
$\tau_{\rm{atm}}(\omega, t)$.
The data that we could not reliably phase connect tended to
involve sites with large wet atmospheric
delays (the St.~Croix VLBA site was generally the most
problematic for phase connection) or be from observations just above
the 10\arcdeg\ elevation threshold.

We used surface meteorological data (barometric pressure, temperature,
and humidity or dew point) along with the equations from
\citet{Saastamoinen+1972} to obtain a priori estimates of
$\tau_{\rm zen\_dry}(t)$ and
$\tau_{\rm zen\_wet}(t)$.
For the mapping functions
$m_{\rm dry}(t, \epsilon)$ and
$m_{\rm wet}(t, \epsilon)$ we used
the formulas provided by
\citet{Niell+1996, Niell+2000}.
Our values for
$\tau_{\rm ion}(\omega, t, \mathbf{\hat s})$
came from
the United States Air Force's
Parameterized Ionosphere Model
\citep[PIM; ][]{Daniell+1995},
which we
adapted for use with VLBI observables
\citep{Campbell+1999}.\footnote[6]{One can obtain
more accurate (we found)
values for 
$\tau_{\rm ion}(\omega, t, \mathbf{\hat s})$
via
the use of
AIPS task TECOR and
publicly available maps of total electron content (TEC) derived
from Global Positioning System (GPS) data
\citep{WalkerC1999}. 
However,
we found no useful TEC maps for our VLBI observations
prior to September 1998. To avoid introducing possible systematic errors
into our astrometric results
from the use of different ionosphere models during different periods of
observations, we opted to use PIM for all of our VLBI
experiments.
Ionosphere model (i.e., PIM) errors contribute insignificantly to the total
standard errors in
the astrometric results for \GPB\@ that we present in Paper~V
\citep{GPB-V}.}

\begin{figure}
\centering
\includegraphics[trim= 0in 3.3in 0in 0in, clip, width=\textwidth]{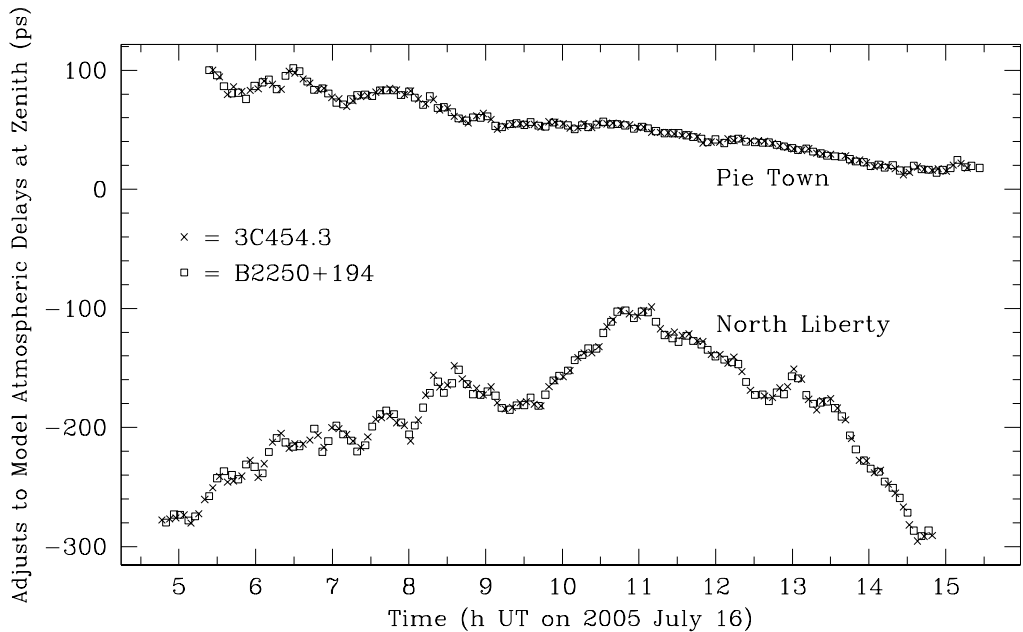}
\caption{Adjustments to model ``wet'' atmospheric delays at zenith
($\tau_{zen\_wet}$)
estimated from the data
with our Kalman filter
for two VLBI sites, Pie Town (NM) and North Liberty (IA), for our
2005 July~16 observations.
For scale, a $2\pi$ radian wrap in fringe phase corresponds to
119~ps of delay along the line of sight
from each antenna
to each observed source.  Although
the adjustments are modeled as wet delays at zenith, they include
contributions from model errors in the ionosphere, too, since those
errors also have elevation-angle dependence.
}
\label{smzenadj}
\end{figure}

In fitting our data to a model, we used a Kalman-filter estimator
\citep[``SOLVK";][]{HerringDS1990} 
that allowed us to model atmospheric delays at zenith
and instrumental drifts as Gauss-Markov stochastic processes.
Our VLBI data do not provide sufficient information
to simultaneously estimate adjustments to
the separate terms
in Equation~\ref{tau_atm};
thus
we adjusted only $\tau_{\rm zen\_wet}(t)$,
the term we assumed to have the least accurate a~priori model.
Figure~\ref{smzenadj} shows the adjustments to
$\tau_{\rm zen\_wet}(t)$ that we obtained with our
Kalman-filter estimator for the data corresponding to
Figures~\ref{figphcon1} and~\ref{figphcon2}.

Through the use of data from two reference sources, \REFA\ and \REFB ,
we were able to distinguish model adjustments to
$\tau_{\rm atm}(\omega, t)$,
which has elevation-angle dependencies, from model adjustments to
other terms, e.g.,
$\tau_{\rm inst}(\omega, t)$, which do not.
Large phase jumps between successive reference-source scans
were generally attributed to atmospheric fluctuations.
Since the size of atmospheric fluctuations roughly scaled with
the signal path length through the atmosphere, phase connection
was generally easiest when the observed sources were at
high elevation angles.  Thus
we generally started our phase-connection process with data obtained during
the middle of an observing session, when the sources were at their highest
elevation angles at antennas near the middle of the array.  We then
worked our way ``outward'' towards the early and late scans of the
session.

When we could not confidently determine the integral
number of $2\pi$ phase wraps between consecutive reference-source
scans, we unweighted the target-source data between those scans.
We also inserted a ``break'' marker at the point of the undetermined
phase jump; the Kalman-filter estimator takes
account of such breaks in its model fit.

\begin{figure}
\centering
\includegraphics[trim= 0in 3.3in 0in 0in, clip, width=\textwidth]{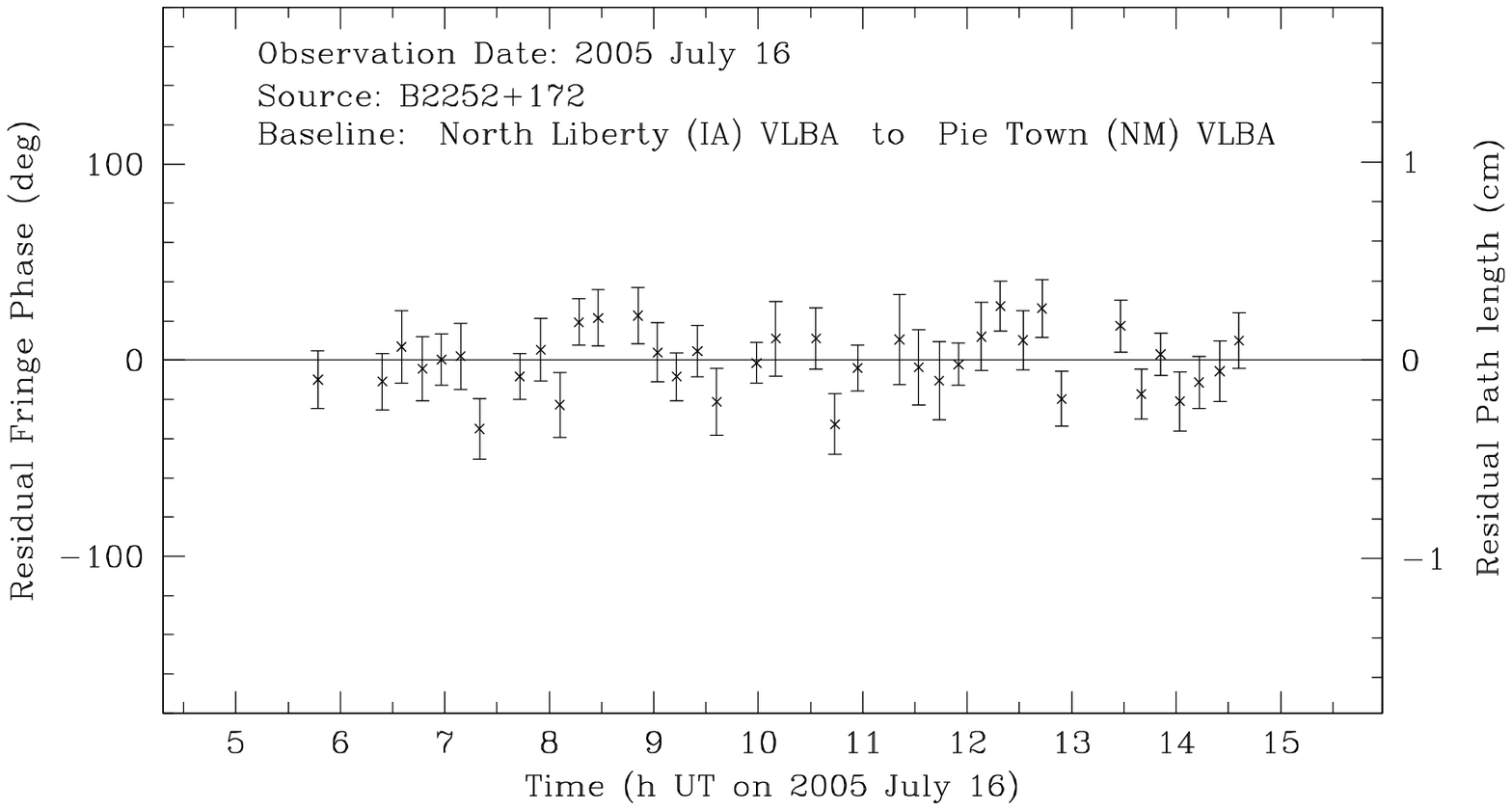}
\caption{Sample phase residuals for source
\protect\REFC.  The phase model was derived
from our reference-source
(\protect\REFA\ and \protect\REFB )
data with our
Kalman-filter estimator.  The residuals shown are for the
North Liberty (Iowa)--Pie Town (New Mexico) baseline and are from
our 2005 July~16 observations.
The short gaps occasionally seen between adjacent data points
are due (in general) either to insufficient signal strength
from \protect\REFC\ to detect the source in a single scan of data, or
to difficulties in determining the integral
number of
$2\pi$ phase wraps between adjacent reference-source data.
The
number of $2\pi$ phase wraps chosen for each point was the number
that minimized the magnitude of the residual.  These magnitudes, and the
differences in phase residuals between adjacent scans, are all
$\ll 2\pi$ radians
($\ll 360\arcdeg$),
which demonstrates that the data are properly phase connected.
We used a relatively conservative estimate for the error of each
phase measurement, so
the scatter of the points relative to the size of the error bars appears
somewhat
small.
}
\label{sm2252res}
\end{figure}

Figure~\ref{sm2252res} illustrates target-source residuals to the
phase models derived from the reference-source data.  The integral
numbers of $2\pi$ phase wraps between successive scans are reliably
determined for all weighted data.

\subsubsection{Final-stage data reduction: use of improved model in
phase-referenced maps}
\label{mergfin}

Once we had phase-delay and fringe-rate models based
on parameter fits to the
reference-source data, we imported those models into
AIPS
to use as the basis for
our phase-referenced
maps of the target source,
either \IMP\ or \REFC .
First we created files of fringe-phase and
fringe-rate
differences between the models derived from the reference-source data and
the models (from the correlator)
originally used in AIPS.  Then we used AIPS task
TBIN to create an AIPS ``solution'' (SN) table with these differences.
We could then treat the change from the original models used in AIPS
to the improved models derived from our reference-source data
as a standard AIPS calibration step.  We
used AIPS task CLCAL to incorporate these model adjustments.  We then
made phase-referenced maps with AIPS task IMAGR following standard
procedures
\citep[e.g.,][]{BeasleyC1995}.
We obtained
positions for our target sources from these IMAGR maps
as described
in Paper~VI
\citep{GPB-VI}.

\section{Comparison of Results}
\label{compare}

We used two 
figures of merit
to compare
the results of our
MA
technique with
the more conventional PRM and PMF techniques:
the quality, as described below, of
the \REFC\ and \IMP\ images produced (MA and PRM techniques only);
and the level of consistency of the astrometric results
for each of these two sources.

\subsection{Comparison of images}

\begin{figure}
\centering
\includegraphics[width=\textwidth]{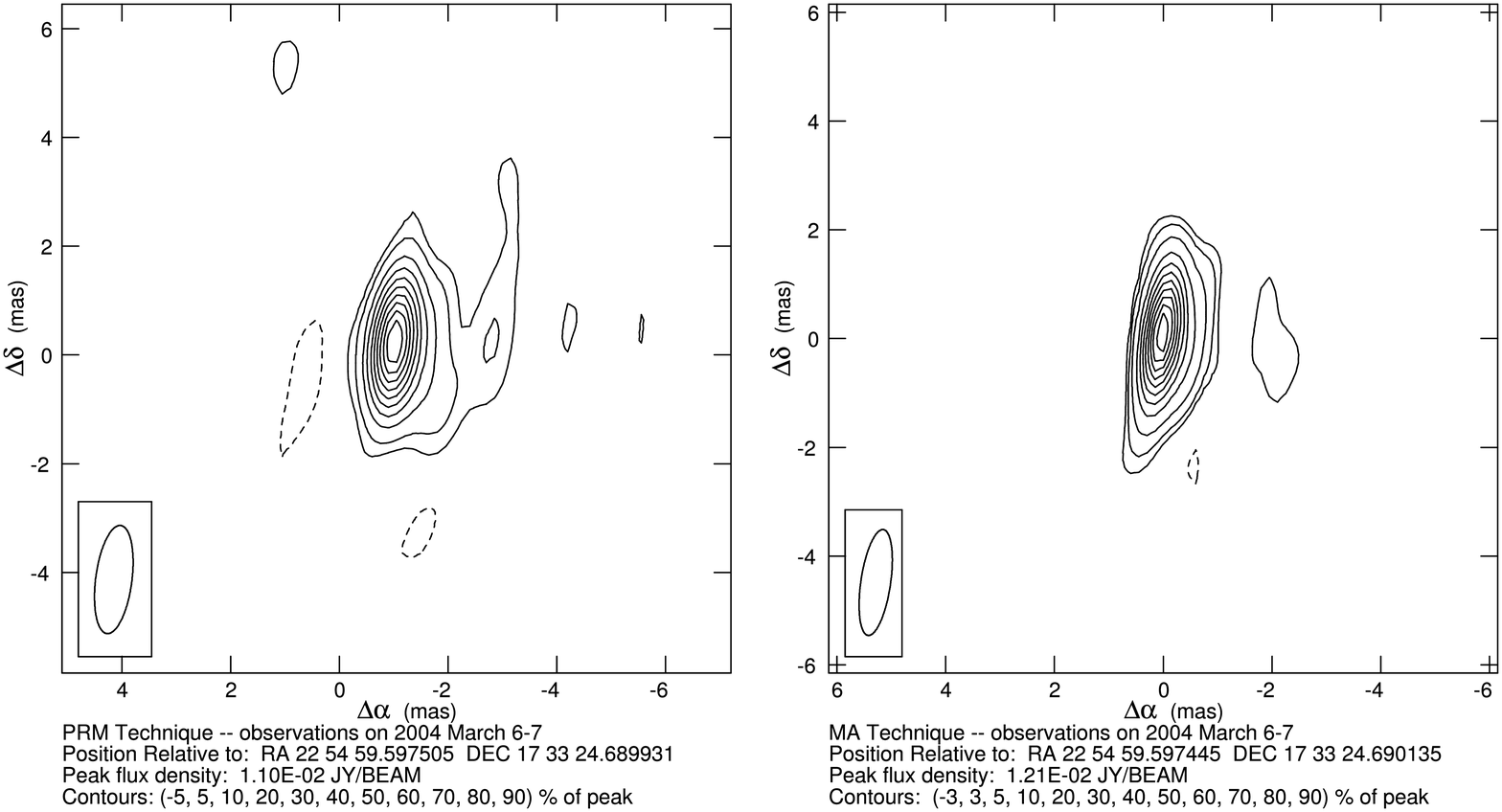}
\caption{Representative images of \protect\REFC\ created
in AIPS with the PRM technique
(left) and the MA technique (right).  The data for the images
are from our observations on 2004 March~6-7.
Note that the two images have different coordinate origins.
For each image, the size and orientation of the restoring beam is
shown in the box in the lower left corner.
The image created with the MA technique is visibly cleaner even with
a lower minimum brightness contour. On average over all of our sessions of
observations, the MA images had 1.9 times the dynamic range of the PRM
images.
}
\label{2252compfig}
\end{figure}

Figure~\ref{2252compfig} shows a representative
map of \REFC\ produced via the conventional PRM technique alongside
the corresponding map generated via our MA technique.
Table~\ref{2252comptab} lists the brightness-peak
flux density, the dynamic range (i.e., the ratio between the
flux density of the brightness peak and the flux density of the
image noise floor),
and
the ratio of
the peak-brightness amplitude ($A_{\rm pk+}$)
to
the amplitude of the most negative peak ($A_{\rm pk-}$)
that we obtained with the PRM and MA techniques
for each of our 12~sessions of observations of this source.
If the extent of the brightness distribution of a source
is comparable to, or larger than, the size of the restoring beam of its image,
then the measured peak brightness and (hence) the dynamic range
are functions of the size of the restoring beam.
The size (and orientation) of the restoring beam
can in turn depend upon which data were weighted (among other factors)
when the image was produced.
Since we employed different data-selection criteria with the PRM and
MA techniques and hence did
not necessarily use identical data
with the two
techniques, we consider
the $A_{\rm pk+}$/$A_{\rm pk-}$ ratio to be the most meaningful measure of image
quality among the three Table~\ref{2252comptab} metrics.

\begin{table}
\begin{center}
\caption{Characteristics of \protect\REFC\ images obtained from the PRM and MA techniques.\label{2252comptab}}
{\scriptsize
\begin{tabular}{c c c c c c c c c c c c c}
\tableline\tableline
& \multicolumn{6}{c}{PRM Technique} & \multicolumn{6}{c}{MA Technique}\\
&\multicolumn{3}{c}{With 10\arcdeg}&\multicolumn{3}{c}{With 30\arcdeg}
&\multicolumn{3}{c}{With 10\arcdeg}&\multicolumn{3}{c}{With 30\arcdeg}\\
&\multicolumn{3}{c}{elev.-angle cutoff} 
&\multicolumn{3}{c}{elev.-angle cutoff}
&\multicolumn{3}{c}{elev.-angle cutoff} 
&\multicolumn{3}{c}{elev.-angle cutoff} \\
Session&&Dyn.&{\multirow{2}{*}{$\frac{A_{\rm pk+}}{A_{\rm pk-}}$}\tablenotemark{c}}&&Dyn.&\multirow{2}{*}{$\frac{A_{\rm pk+}}{A_{\rm pk-}}$}&
&Dyn.&\multirow{2}{*}{$\frac{A_{\rm pk+}}{A_{\rm pk-}}$}&
&Dyn.&\multirow{2}{*}{$\frac{A_{\rm pk+}}{A_{\rm pk-}}$}\\
Start Date&$A_{\rm pk+}$\tablenotemark{a}&Range\tablenotemark{b}&
&$A_{\rm pk+}$&Range&&$A_{\rm pk+}$&Range&&$A_{\rm pk+}$&Range&\\
\tableline
2002-11-20 & 12.2 & 64 & \phantom{0}9.7 & 14.2 & \phantom{0}68 & 13.\phantom{0} & 15.2 & 165 & 37.\phantom{0} & 15.5 & 170 & 31.\phantom{0} \\
2003-01-26 & 11.3 & 53 & \phantom{0}5.6 & 14.1 & \phantom{0}70 & \phantom{0}8.3 & 17.0 & 140 & 29.\phantom{0} & 17.9 & 149 & 38.\phantom{0} \\
2003-05-18 & 12.5 & 62 & \phantom{0}7.8 & 16.6 & \phantom{0}60 &  \phantom{0}8.7 & 19.1 & 187 & 47.\phantom{0} & 19.8 & 173 & 44.\phantom{0} \\
2003-09-08 & 11.7 & 64 & 13.\phantom{0} & 13.7 & \phantom{0}72 & 16.\phantom{0} & 15.4 & 141 & 34.\phantom{0} & 16.0 & 132 & 34.\phantom{0} \\
2003-12-05 & 11.0 & 51 & \phantom{0}7.2 & 14.3 & \phantom{0}61 &  \phantom{0}9.9 & 15.4 & 106 & 16.\phantom{0} & 16.0 & \phantom{0}97 & 20.\phantom{0} \\
2004-03-06 & \phantom{0}8.9 & 56 & \phantom{0}8.8 & 11.0 & \phantom{0}76 & 12.\phantom{0} & 12.1 & 124 & 30.\phantom{0} & 12.4 & 110 & 25.\phantom{0} \\
2004-05-18 & \phantom{0}6.5 & 37 & \phantom{0}7.3 & \phantom{0}8.8 & \phantom{0}48 & \phantom{0}8.8 & 10.0 & \phantom{0}77 & 18.\phantom{0} & 10.5 & \phantom{0}68 & 18.\phantom{0} \\
2004-06-26 & \phantom{0}7.0 & 54 & 11.\phantom{0} & \phantom{0}9.0 & \phantom{0}70 & 14.\phantom{0} & \phantom{0}9.8 & 109 & 25.\phantom{0} & 10.1 & 103 & 22.\phantom{0} \\
2004-12-11 & \phantom{0}9.0 & 58 & \phantom{0}8.9 & 12.0 & \phantom{0}94 & 14.\phantom{0} & 12.2 & 128 & 34.\phantom{0} & 12.8 & 116 & 30.\phantom{0} \\
2005-01-15 & \phantom{0}9.7 & 87 & 12.\phantom{0} & 11.0 & 103 & 12.\phantom{0} & 11.3 & 175 & 39.\phantom{0} & 11.6 & 149 & 38.\phantom{0} \\
2005-05-28 & 11.7 & 48 & 10.0 & 15.5 & \phantom{0}69 & 15.\phantom{0} & 16.3 & 125 & 36.\phantom{0} & 16.5 & 114 & 33.\phantom{0} \\
2005-07-16 & 12.5 & 64 & 11.\phantom{0} & 15.4 & \phantom{0}80 & 14.\phantom{0} & 16.0 & 127 & 27.\phantom{0} & 16.5 & 131 & 28.\phantom{0} \\
\tableline
 MEAN\tablenotemark{d} & 10.3 &  58.1 &  \phantom{0}9.3 & 13.0 &  72.6 &  12.0 & 14.2 & 133.6 &  31.0 & 14.6 & 125.9 &  30.0 \\
\tableline
\end{tabular}
} 
\tablenotetext{a}{Brightness-peak amplitude, in mJy/beam.}
\tablenotetext{b}{Dynamic range (see text).}
\tablenotetext{c}{Ratio of the brightness-peak amplitude ($A_{\rm pk+}$) to the amplitude of the most negative peak ($A_{\rm pk-}$).}
\tablenotetext{d}{Unweighted mean over all sessions.}
\end{center}
\end{table}

For both the PRM and MA techniques,
we show these metrics for images 
produced after we applied either 
a 10\arcdeg\ or a 30\arcdeg\ elevation-angle cutoff to the data
(i.e., after we unweighted all data from each site whenever the observed
source was below the specified elevation angle).
Since the major source of phase-calibration
error with the PRM technique is often
poor modeling of
the atmosphere, and since (as mentioned earlier)
atmospheric model errors tend to get
larger at low elevation angles,
the removal of low-elevation-angle
data can potentially improve the quality of both the image
and the astrometry obtained
with the PRM technique.
(Note that, for the PRM technique,
the images when a 30\arcdeg\ elevation-angle cutoff is used nearly always have
higher dynamic ranges and $A_{\rm pk+}$/$A_{\rm pk-}$ ratios than the corresponding images when
a 10\arcdeg\ elevation-angle cutoff is used.)
For the MA technique,
only those
time spans with reliably phase-connected data are weighted,
so we might expect a degradation of image quality
(due to the reduction of weighted data) when a higher elevation-angle
cutoff is imposed.
The results found for our images are mostly consistent with this
expectation, although Table~\ref{2252comptab} shows
that for \REFC, the higher elevation-angle cutoff did improve the $A_{\rm pk+}$/$A_{\rm pk-}$
ratio with the MA technique in four of the twelve sessions.
In any event,
even if, for the PRM technique, we select for each session the
elevation-angle cutoff
in Table~\ref{2252comptab} with the highest dynamic range or $A_{\rm pk+}$/$A_{\rm pk-}$
ratio for each session,
the corresponding MA images
for a 10\arcdeg\ elevation-angle cutoff have,
on average, 1.9~times the dynamic
range and 2.7~times the $A_{\rm pk+}$/$A_{\rm pk-}$ ratio of the PRM images.
The MA images, when compared with the PRM images,
also show significantly less
variability in peak flux density for the two different elevation-angle
cutoffs that are used.
This superior consistency, too, suggests that the
MA technique provides more accurate model calibration than the
PRM technique.

\begin{figure}
\centering
\includegraphics[width=\textwidth]{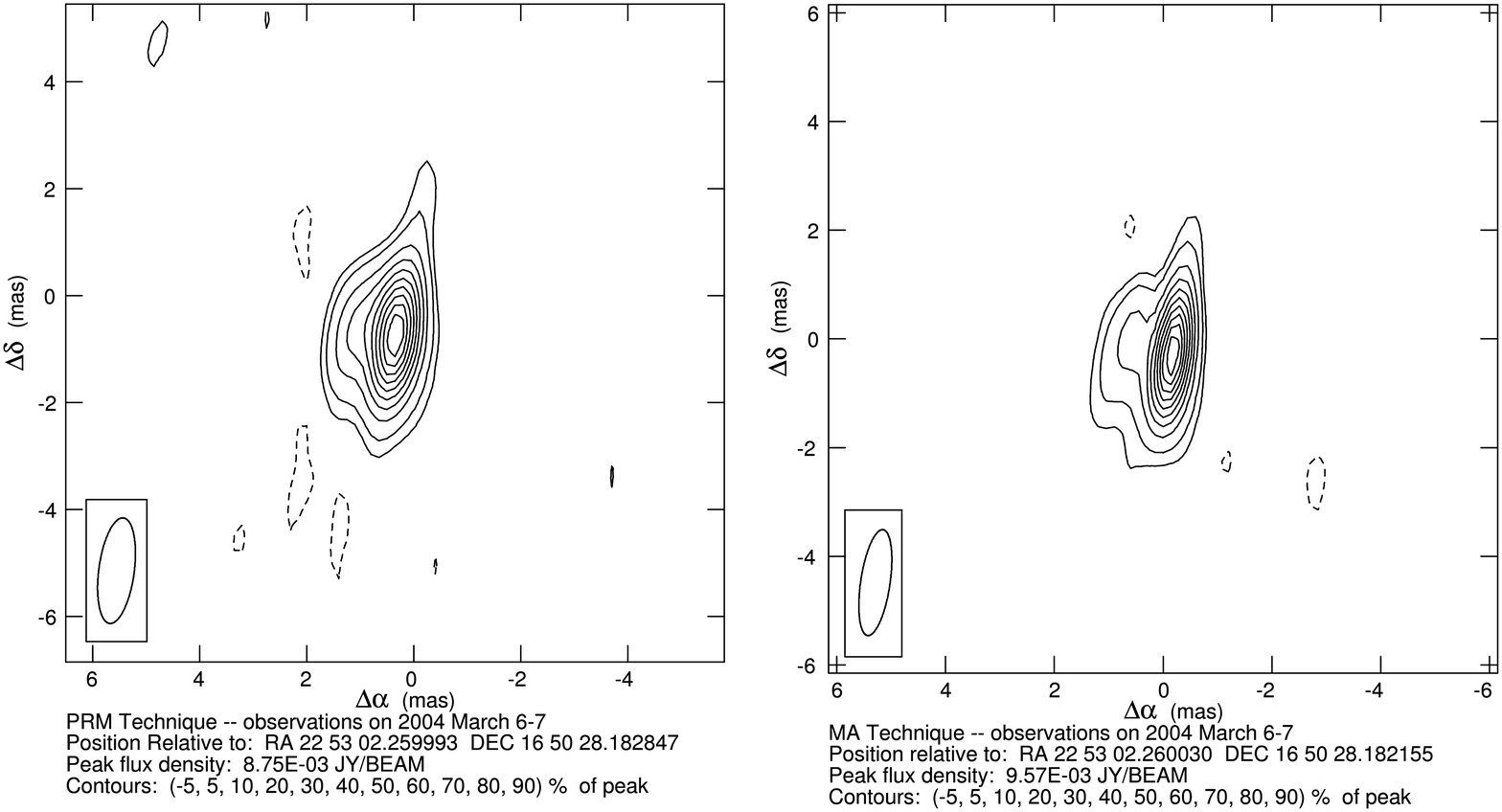}
\caption{Representative images of \protect\IMP\ created in AIPS
with the
PRM technique
(left) and the MA technique (right), from
observations on 2004 March~6-7.
For each image, the size and orientation of the restoring beam are
shown in the box in the lower left corner.
Note that the left and right images have different coordinate
origins.
}
\label{hr8703compfig}
\end{figure}

Figure \ref{hr8703compfig} shows representative images of \IMP\ produced
with both the PRM and MA techniques.
Averaged across all of our observing sessions,
the peak flux density we detected for \IMP\ was
16\% higher with the MA technique than with the
PRM technique.
However, the significance of this result is unclear,
because we used different
restoring beams for each technique.
Unfortunately,
the data files used to make the images
of \IMP\ with the PRM technique
were no longer available
at the time of preparation of this publication,
so we could not
use restoring beams with the same size and orientation for both techniques.
We also did not
have noise-floor or ``minimum peak'' information
available
for these PRM images.
In addition, we know that \IMP\ can vary in
total brightness
and brightness distribution (i.e., radio structure)
over hour time scales 
\citep[Paper VII,][]{GPB-VII, Lebach+1999},
which further complicates comparisons
between the techniques, especially since
a constant flux density over the
full duration of a session of observations is inherently assumed
in the imaging process for both techniques. 
Thus, for comparison of image quality from the PRM and MA techniques,
we considered the \REFC\ results a better measure than the \IMP\ results. 

\subsection{Comparison of astrometric results}
\begin{figure}
\centering
\includegraphics[width=0.85\textwidth]{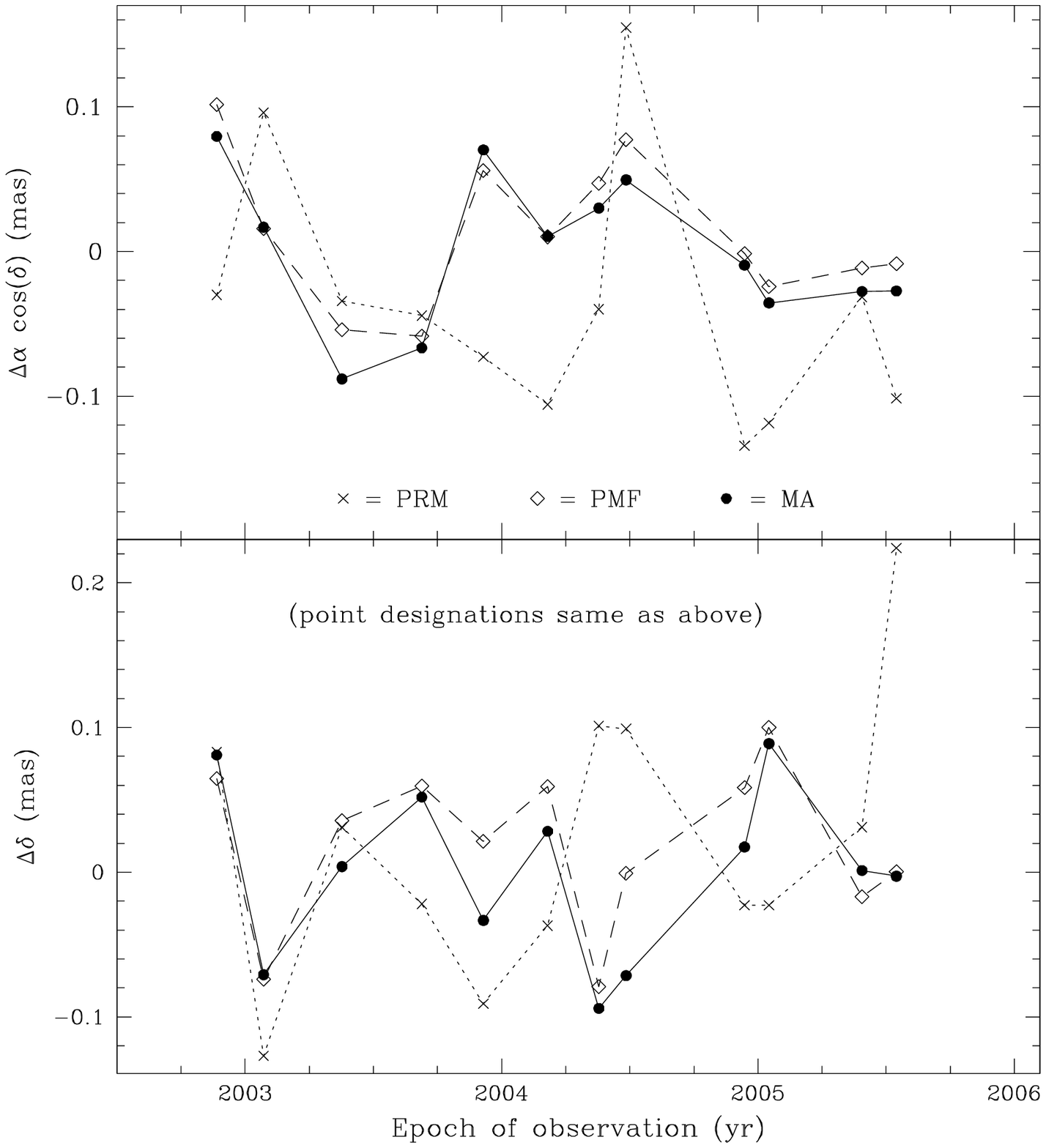}
\caption{Position estimates for
\protect\REFC\ relative to \protect\REFA\ 
from the PRM (diagonal crosses; connected by dotted line
segments), PMF (diamonds; connected by dashed line segments),
and MA (filled circles; connected by solid line segments) techniques (see text).
The zero position for each 
coordinate is the unweighted mean value
of the twelve estimates obtained with the MA technique.
For better plot clarity, the error bars for the plotted
points are not shown.  Since
the scatter in the position estimates is
due mainly to
model errors
(such as for the ionosphere)
rather than to SNR limits of the data,
the rms scatter
of the position estimates
relative to the mean
for each technique,
as given in Table~\ref{2252postab}, is a
good approximation for the errors of the corresponding points shown in
this plot.  Such errors ignore possible systematic effects that
could have shifted the mean positions for each of the three techniques.
}
\label{2252posplt}
\end{figure}

Figure \ref{2252posplt} compares the
position estimates for
\REFC\ (relative to \REFA ) from the
PRM, PMF, and
MA techniques.
(We used a 30\arcdeg, rather than 10\arcdeg , elevation-angle cutoff
of data with the PRM technique, because the higher cutoff yielded
better results for this technique.
We used our nominal
10\arcdeg\ elevation-angle cutoff of data with the PMF and MA techniques.)
Table~\ref{2252postab} shows the root-mean-square
(rms) scatter
about the mean position for each technique.
When, as with \REFC ,
the target source is commonly detected on
many baselines in a single scan
and structure corrections with the PMF technique are available,
the MA technique shows no astrometric advantage over the PMF technique.
However,
our MA technique yields
position estimates with only about
60\% of the rms scatter of the estimates from the PRM technique,
i.e.,
removes
the equivalent of a noise source of over 0.06~mas per
coordinate estimate associated with the PRM technique.

\begin{table}
\begin{center}
\caption{Astrometric results for \protect\REFC\ from
the PRM, PMF, and MA techniques.\label{2252postab}}
\begin{tabular}{l@{\hspace{30pt}}c c@{\hspace{40pt}}c c@{\hspace{40pt}}c c}
\tableline\tableline
& \multicolumn{2}{c}{Mean Offset\tablenotemark{a}} &
\multicolumn{2}{c}{RMS Scatter\tablenotemark{b}} &
\multicolumn{2}{c}{Standard Error} \\
& \multicolumn{2}{c}{(mas)} &
\multicolumn{2}{c}{(mas)} &
\multicolumn{2}{c}{of mean\tablenotemark{c} (mas)} \\
Technique & \multicolumn{1}{c}{$\alpha$} & \multicolumn{1}{c}{$\delta$} &
$\alpha$ & $\delta$ & $\alpha$ & $\delta$ \\
\tableline
PRM & $-$0.039 & 0.021 & 0.082 & 0.092 & 0.025 & 0.028 \\
PMF & \phantom{$-$}0.012 & 0.019 & 0.048 & 0.053 & 0.014 & 0.016 \\
MA  & \phantom{$-$}0.000 & 0.000 & 0.050 & 0.057 & 0.014 & 0.017 \\
\tableline
\end{tabular}
\tablenotetext{a}{Defined to be zero for the MA technique
in both right ascension ($\alpha$) and declination ($\delta$).
The mean offsets for the PRM and PMF techniques
are relative to the mean coordinate position
from the MA technique.}
\tablenotetext{b}{Relative to the
mean coordinate position for each technique.}
\tablenotetext{c}{Calculated from the rms scatter.}
\tablecomments{For each technique we assumed, for each of $\alpha$ and $\delta$,
the same
standard error for
every
observing session.
The rms scatter for the PRM technique is the largest;
the levels of rms scatter
for the PMF and MA techniques are not significantly different
from each other, although the former appear to be slightly smaller.}
\end{center}
\end{table}

The model for \IMP's position over time must include proper motion,
parallax, and
orbital motions associated with the binary system.
We used the preferred parametrization of Paper~V
\citep{GPB-V},
one in which no proper acceleration is estimated,
to obtain the results presented here.

For seven of the 35~observing sessions of \IMP, including
the most astrometrically important final three,
the radio emissions
from the star were insufficiently bright
to obtain reliable detections of \IMP\ in single scans.
Thus we could not obtain position
estimates of \IMP\ with the PMF technique for any of these sessions.
We also found from images obtained with the PRM and
MA techniques
that \IMP\ had relatively complex structure during several of
our observing sessions
\citepalias[see][]{GPB-VII},
and the PMF technique by
itself cannot well identify or characterize
such structure.
Thus we deemed the PMF technique unsuitable
for astrometric studies of \IMP\ with our data.

\begin{figure}
\centering
\includegraphics[width=0.85\textwidth]{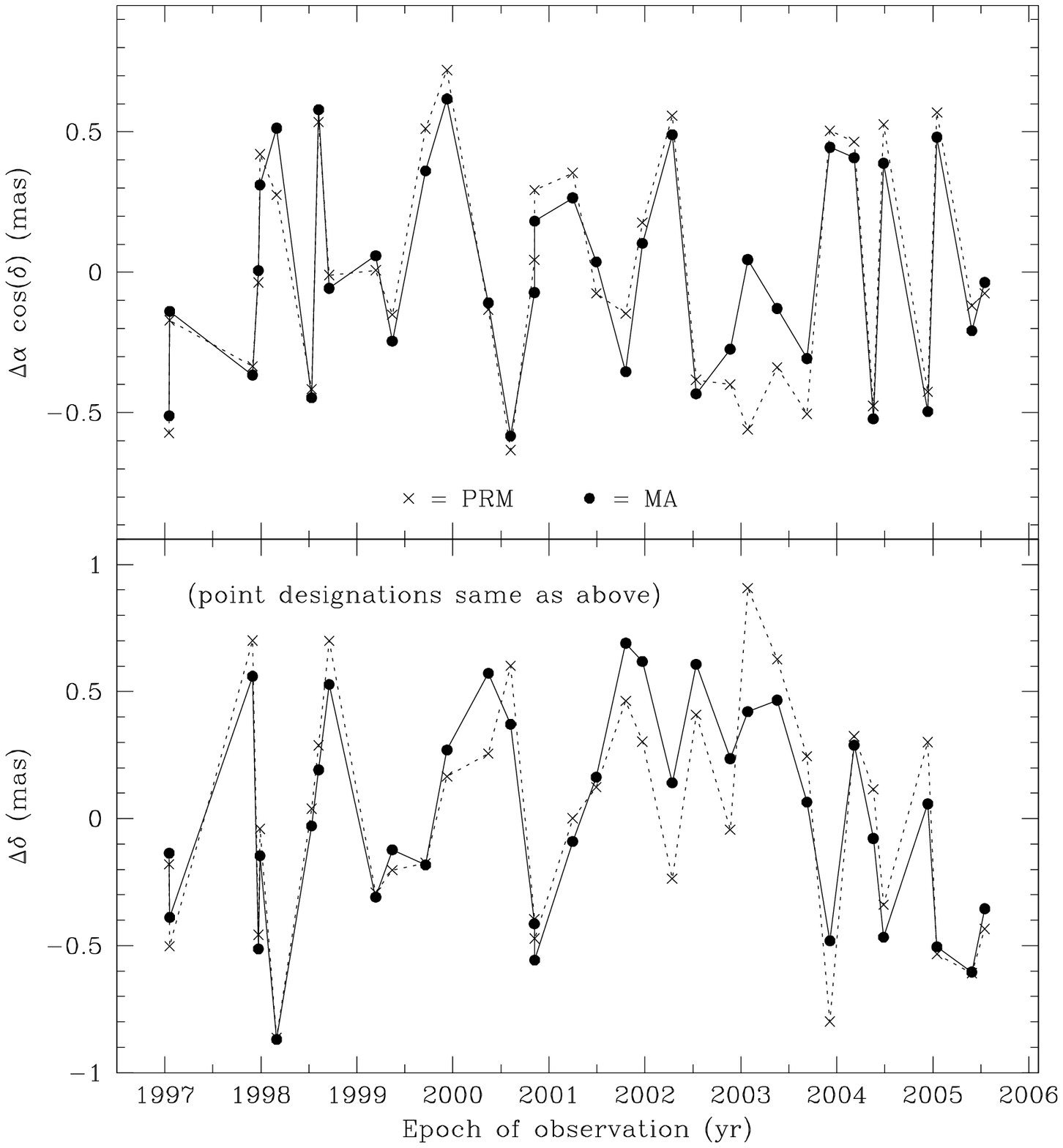} 
\caption{Postfit residuals of the
position estimates of
\protect\IMP\ from the PRM (diagonal crosses; connected by dotted line
segments) and
MA (filled circles; connected by solid line segments) techniques.
(In the model fit, we estimate the 
position, proper motion, parallax, and apparent binary-system
orbit of the stellar 
radio emitting region.)
We could not obtain position estimates of \protect\IMP\ with the
PMF technique
for seven of our 35 observing sessions due to insufficient radio brightness
of the star; hence we deemed the standard PMF technique
unsuitable for
astrometric studies of \protect\IMP\ with our data.
For better plot clarity, the error bars for the plotted
points are not shown.
The scatter in the position estimates from both the
PRM and MA techniques is
due mainly to
intrinsic motions of the radio emission of \protect\IMP\ relative to the
primary component of the star
rather than to errors
in the position estimates of the radio emission.
}
\label{hr8703posplt}
\end{figure}

In Figure~\ref{hr8703posplt} we show the post-fit residuals of the
position estimates for
\IMP\ from
the PRM and MA techniques, and in
Table~\ref{hr8703postab} the rms scatter of these same pos-tfit residuals.
On average the post-fit residuals from
our MA technique
had about 10\% less scatter in
right ascension and 7\% less scatter in declination than the
corresponding residuals from the PRM technique.
The scatter
from both techniques is likely dominated by the
unmodeled intrinsic motions of the stellar radio emission relative to the 
star's primary component \citepalias{GPB-VI}.
Hence,
since we have no model
that can accurately describe the motion of the
radio emission of \IMP\ relative to its primary,
we can expect at most only
a minor improvement
in the rms scatter of position residuals
from the use of any superior astrometric technique.
Indeed the MA technique yielded just such a minor improvement.

\begin{table}
\centering
\caption{RMS scatter of
\protect\IMP\ position residuals
obtained from the PRM and MA techniques.\label{hr8703postab}}
\begin{tabular}{l@{~~~~~}c c}
\hline\hline
& \multicolumn{2}{c}{RMS Scatter} \\
& \multicolumn{2}{c}{(mas)} \\
Technique & $\alpha$ & $\delta$ \\
\hline
PRM & 0.40 & 0.44 \\
MA & 0.35 & 0.41 \\
\hline
\end{tabular}
\tablecomments{The position residuals
are relative to an
astrometric model fit in which
the radio emissions
are assumed to be tied to a single stellar component of
the \protect\IMP\ binary system.
We believe that the rms scatter is due predominantly to intrinsic motions
of the radio emissions relative to this stellar component rather than
to measurement accuracy.  Nevertheless, the rms scatter is
roughly 10\% smaller with
the MA technique
than with the PRM technique.}
\end{table}

\section{Conclusions}
\label{conclusion}

In comparing the astrometric results from phase-referenced mapping
(PRM), parametric model fitting (PMF), and the merged analysis (MA)
technique introduced
in this paper, we found:
\begin{enumerate}
\item The PRM technique
can
be used with relatively weak radio sources.
It is the least labor intensive
among the techniques tested,
but it also provides the least astrometric accuracy.
(With this technique, unlike with the PMF or MA techniques,
we were unable to use
multiple reference sources to better account for
the elevation-angle dependence of
model errors; such use
could improve the astrometric accuracy attainable with this technique.)
\item The PMF technique is useful only when both reference and target
sources are: (i)~reliably detectable in short (i.e., few-minute) scans of
data; and (ii)~pointlike, unless structure corrections are available from
other means.
Under these circumstances, the PMF technique yielded the highest
astrometric accuracy among the techniques tested.
\item Our MA technique can provide astrometric accuracies
nearly equal to those obtained with the PMF technique.
\item Very importantly, our MA technique can be used with sources that are too
weak for use
of the PMF technique,
as was \IMP\ during seven of our 35 VLBI sessions.  
In general, any source that is sufficiently bright for 
use of the PRM technique is sufficiently bright for use of our MA technique.
The MA technique can also be used with
sources that have complex brightness distributions and was shown to yield
images with significantly higher dynamic ranges than comparable images from
the PRM technique.
\end{enumerate}
Our use of the MA technique enabled us to obtain more accurate
astrometric measurements
of \IMP , the guide star for the \GPB\ mission,
than we could have otherwise obtained with the conventional
PRM or PMF techniques.

\acknowledgements
ACKNOWLEDGMENTS. We thank N. Nunes
for developing the graphical
software package we used to phase connect our VLBI data.
We thank the VLBI group of NASA's Goddard Space Flight Center,
L.~Petrov in particular, for providing technical support and information.
We are also grateful to the many people involved in our campaign
of VLBI observations who went above and beyond
the call of duty, including S.~Dains,
C.~Garcia Miro, E.~Moll, and L.~Cameron.
We thank C.~Jacobs and O.~Sovers for information about the coordinates
of the three 70~m NASA DSN antennas used in our observations, and
R.~C.~Walker for information about the VLA coordinates.
We thank J.~L.~Davis and S.~S.~Shapiro for their 
information and insights on the use of our
Kalman-filter estimator (SOLVK).
Finally, we thank R.~C.~Walker, K.~Desai, and E.~Greisen for their support
in our use of AIPS.
The National Radio Astronomy Observatory
(NRAO) is a facility of the National Science Foundation operated under
cooperative agreement by Associated Universities, Inc.
This research made use of NASA's Astrophysics Data
System, which was
conceived, developed, and continues to be
operated by the Smithsonian Astrophysical
Observatory at the Harvard-Smithsonian Center for Astrophysics.
Our work
was supported by NASA prime award NAS8-39225, Stanford
University sub-award PR\thinspace 6750,
the Smithsonian Institution, and York University.

\end{document}